\documentclass[12pt,preprint]{aastex}
\usepackage{emulateapj5,natbib,multicol}
\begin{document}
\def\SZ{Sunyaev-Zel'dovich~}
\def\ie{i.e.}
\def\eg{e.g.}
\def\etal{et al.}
\def\hou{km s$^{-1}$ Mpc $^{-1}$}
\def\omega0{$\Omega_\circ$}
\def\omegam{$\Omega_M$}
\def\omegal{$\Omega_\Lambda$}
\def\lambdao{$\Lambda_\circ$}
\def\Ho{$H_{\circ}$}
\def\rc{$\theta_c$}
\def\be{$\beta$}
\def\dT{$\Delta T(0)$}
\def\rms{$r.m.s.$}
\def\uv{$u$-$v$}
\def\r500{$r_{500}$}
\def\dt{$\Delta T(0)$}
\def\dchi{$\Delta\chi^2$}
\def\qo{$q_\circ$}
\def\da{$D_A$}
\def\rosat{{\em ROSAT}}
\def\chandra{{\em Chandra}}
\def\xmm{{\em XMM}}
\def\hectospec{{\em HECTOSPEC}}
\def\da{$D_A$}
\def\cf{{\em cf.}}
\def\LameH{{$\Lambda_H$}}
\def\muH{{$\mu_H$}}
\def\ms1137{MS\,1137.5+6625}
\def\chisq{$\chi^2$}
% Journal abbreviations (SPW)
%
\newcommand{\vol}[1]{{\bf{ #1}}}
\newcommand{\nature}{\em Nature\rm}
\newcommand{\aanda}{{\em A. \& A.{\rm}}}
%
%X-ray Missions, Instruments
%
\newcommand{\ein}{{\em Einstein{\rm}}}
\newcommand{\exo}{{\em EXOSAT{\rm}}}
\newcommand{\ginga}{{\em Ginga{\rm}}}
\newcommand{\spek}{SPEKTROSAT}
\newcommand{\asca}{{\em ASCA}}
\newcommand{\free}{Freestanding Instrument}
\newcommand{\hxt}{hard x-ray telescope}
\newcommand{\slew}{Slew Survey}
%
%X-ray Terminology
%
\newcommand{\logn}{{\em logN-logS{\rm}}}
\newcommand{\xrb}{x-ray background}
\newcommand{\dxrb}{diffuse x-ray background}
\newcommand{\aox}{\ifmmode{\alpha_{\tiny OX}} \else $\alpha_{\scriptsize OX}$\fi} 
\newcommand{\alpe}{\ifmmode{\alpha} \else $\alpha$\fi}
\newcommand{\atoms}{\ifmmode{\rm ~atoms~cm^{-2}} \else ~atoms cm$^{-2}$\fi}
\newcommand{\nh}{\ifmmode{\rm N_{H}} \else N$_{H}$\fi}
%\newcommand{\20}{\ifmmode \times 10^{20}\else$\times 10^{20}$\fi}
%
%Non-X-ray Telescopes
%
\newcommand{\kao}{Kuiper Airborne Observatory}
\newcommand{\iue}{\em International Ultraviolet Explorer}
%
%Science
%
\newcommand{\qed}{quasar energy distribution}
\newcommand{\qeds}{quasar energy distributions}
\newcommand{\ir}{infrared}
\newcommand{\optuv}{optical/ultraviolet}
\newcommand{\nufnu}{\ifmmode \nu f_{\nu} \else$\nu f_{\nu}$\fi}
\newcommand{\fnu}{\ifmmode f_{\nu} \else$f_{\nu}$\fi}
\newcommand{\sn}{signal-to-noise}
\newcommand{\ede}{E/$\Delta$E}
\newcommand{\agn}{active galactic nuclei}
\newcommand{\snr}{supernova remnants}
\newcommand{\mdot}{$\dot{M}$}
\newcommand{\msun}{$M_{\odot}$}
\newcommand{\rr}{$F_{PL} \over F_{disk}$}
\newcommand{\costheta}{$\cos \theta$}
\newcommand{\astar}{$a_{*}$}
\newcommand{\bapp}{\ifmmode \beta_{app}\else$\beta_{app}$\ \fi}
% AASTeX Macros for those cases in which AASTex cannot be used
%
%\newcommand{\ion}[2]{#1$\;${\small\rm\@Roman{#2}}\relax}
%\newcommand{\nodata}{ ~$\cdots$~ }
% Units
%
%\newcommand{\micron}{\ifmmode\mu{\rm m}\else$\mu{\rm m}$\fi}
\newcommand{\erg}{\ifmmode erg~cm^{-2}~s^{-1}\else erg cm$^{-2}$ s$^{-1}$\fi}
\newcommand{\degs}{\ifmmode ^{\circ}\else$^{\circ}$\fi}
\newcommand{\pmn}{\ifmmode \pm\else$\pm$\fi}
\newcommand{\per}[1]{\ifmmode \^{#1}\else$^{#1}$\fi}
\def\farcm{\hbox{$.\mkern-4mu^\prime$}}
\def\farcs{\hbox{$.\!\!^{\prime\prime}$}}
\def\fdg{\hbox{$.\!\!^\circ$}}
%
% English abbreviations. add to taste.
%
%\newcommand{\ie}{{\em i.e.}}
\newcommand{\vs}{{\em vs.}}
\newcommand{\via}{{\em via}}
\newcommand{\esp}{{\em esp.}}
\newcommand{\etc}{{\em etc.}}
\newcommand{\twid}{\ifmmode\sim \else${\sim}$\fi}
\newcommand{\lapprox}{_<\atop^\sim}  % math mode only!
\newcommand{\gapprox}{_>\atop^\sim}  % math mode only!
\newcommand{\my}{mas~yr$^{-1}$ }

\include{psfig}

\title{A Deep Chandra Observation of the Distant Galaxy Cluster MS1137.5+6625}

\author{Laura Grego\altaffilmark{1}, Jan Vrtilek\altaffilmark{1}, Leon
Van Speybroeck\altaffilmark{1},  Laurence P. David\altaffilmark{1},
William Forman\altaffilmark{1}, John E. Carlstrom\altaffilmark{2},
Erik D. Reese\altaffilmark{3}, \and Marshall
K. Joy\altaffilmark{4}}

\email{lgrego@head-cfa.harvard.edu, jvrtilek@head-cfa.harvard.edu,
lvanspeybroeck@cfa.harvard.edu, jc@hyde.uchicago.edu,
reese@cfpa.berkeley.edu, marshall.joy@msfc.nasa.gov,
ldavid@head-cfa.harvard.edu, wforman@head-cfa.harvard.edu}
\altaffiltext{1}{Harvard-Smithsonian Center for Astrophysics, 60 Garden 
St, Cambridge, MA 02138}
\altaffiltext{2}{Department of Astronomy \& Astrophysics, 5640 S. Ellis 
Ave., University of Chicago, Chicago, IL 60637}
\altaffiltext{3}{Department of Physics, University of California, 
Berkeley, CA 94720, Chandra Fellow}
\authoraddr{Harvard-Smithsonian Center for Astrophysics, MS 83, 60 
Garden St., Cambridge, MA 02139}
\altaffiltext{4}{Dept. of Space Science, SD50, NASA Marshall Space 
Flight Center, Huntsville, AL 35812}
%\altaffiltext{4}{Department of Physics, University of Alabama, 
%Huntsville, AL 35899}

\begin{abstract}
We present results from a deep \chandra\ observation of MS1137.5+66, a
 distant (z=0.783) and massive cluster  of galaxies.  
 Only a few similarly massive clusters are currently known
 at such high redshifts; accordingly, this observation provides
 much-needed information on the dynamical state of these rare systems.
The cluster appears both regular and symmetric in the X-ray 
image. However, our analysis of the spectral and spatial X-ray data in conjunction
 with interferometric Sunyaev-Zel'dovich effect data and published
 deep optical imaging suggests the cluster has a fairly complex
 structure. The angular diameter distance we calculate from the \chandra\ and Sunyaev-Zel'dovich effect data assuming an isothermal, spherically symmetric cluster implies a low value for the Hubble constant for which we explore possible explanations.  
\end{abstract}

\section{Introduction}
\label{sec:intro}

Rich galaxy clusters are useful cosmological probes.  The
richest clusters are thought to be massive enough to comprise a
fair sample of the Universe, \ie, their mass composition should reflect 
the universal mass composition
\citep{white1993,evrard1997} and so provide an efficient laboratory 
for
measuring the baryonic to dark matter ratio.  
The mere existence of massive galaxy clusters at high redshifts can 
also place powerful
constraints on the physical and cosmological parameters of structure
formation models 
(e.g., \citealt{peebles1989,bahcall1992,luppino1995,oukbir1997,bahcall1998,donahue1998,eke1998,haiman2001,holder2001}). The greatest leverage is provided by 
the
most massive and distant clusters;
in fact, to constrain $\Omega_\Lambda$ one must use clusters with
redshifts greater than 0.5  (e.g., \citealt{oukbir1997,holder2001}).  For these reasons, we are particularly
interested in observations of massive galaxy clusters at high redshift.

A set of observations of distant clusters can be used to constrain the
expansion rate and curvature of the universe.  The most noted approach
is through the angular diameter distance relation.  The
theoretical value of the angular diameter distance $D_A(z)$ is 
cosmology-dependent; $D_A(z)$ can be calculated from measurements of the X-ray
temperature and surface brightness profile of a cluster's intracluster
medium (ICM) analyzed in
conjunction with a measurement of the cluster's Sunyaev-Zel'dovich
effect 
\citep[\cf][]{birkinshaw1991,birkinshaw1994,myers1997,hughes1998,reese2000a,patel2000,grainge2002}, a spectral
distortion of the cosmic microwave background (CMB) radiation by the
hot, ionized ICM \citep{sunyaev1970,sunyaev1972}.

Another approach to measuring cosmological constants with clusters
uses the idea that the fraction of the total cluster mass contained in this atmosphere, 
the gas mass fraction $f_g$,
traces the universal baryonic mass fraction, under the fair sample
assumption. The fair sample assumption then implies also that
the gas mass fraction in massive clusters should not vary.

With a well-chosen sample which includes clusters at high-redshifts,
both the above determinations of the cluster gas mass fraction
\citep[\cf][]{sasaki1996,pen1997,rines1999} and cluster distances can 
 be used to constrain the
geometry of the universe, since at high redshifts, the calculated
cluster distances and gas mass fractions depend significantly on the 
cosmological
parameters \omegam\ and \omegal.
 Recent measurements of $D_A(z)$ via the X-ray/SZE method \citep{mason2001,reese2002,jones2003}
and of the gas mass fraction \citep{grego2001} in samples of clusters 
imply values
of the Hubble constant and the curvature of the universe in good
agreement with the values determined independently with other methods.
 In the work of \cite{reese2002}, the angular diameter distance to 18
clusters, distributed widely in redshift, is calculated.  For \omegam=0.3,
\omegal=0.7, they find the mean Hubble constant for the sample is
60$^{+4}_{-4} \ ^{+18}_{-13}$ km s$^{-1}$ Mpc$^{-1}$.  For seven
low-redshift clusters, \cite{mason2001} find a mean Hubble constant of
66$^{+14}_{-11}\pm$15  km s$^{-1}$ Mpc$^{-1}$ for this same 
cosmology. \cite{jones2003} find a mean value of 65$^{+8}_{-7}\pm$15 km
s$^{-1}$ Mpc$^{-1}$ for five moderate-redshift clusters. (The
uncertainties, statistical and systematic, are reported at 68\%
confidence.)  These Hubble constant measurements agree within the
uncertainties with the value derived from the Hubble Space Telescope
\Ho\ Key Project \cite[\cf][]{mould2000}, 72$\pm3\pm7$ km s$^{-1}$ Mpc$^{-1}$, and Type Ia supernovae
\cite[\cf][]{riess1998}, who find the Hubble constant to be $\sim$65,
in the cosmology considered here.
 
When determining cosmological parameters from cluster distances and
mass fractions, one generally assumes the dominant cluster physics
is well-represented by a simple picture: galaxies and an ionized
intra-cluster medium in hydrostatic equilibrium in a large dark-matter
potential, supported by thermal pressure.  In general, this is
descriptive of clusters in the local universe, as evidenced by the 
strong correlations between X-ray luminosity and temperature \citep{mushotzky1997,allen1998,arnaud1999,donahue1999} and cluster size
and temperature \citep{mohr2000}; and by the
uniformity of the gas mass fraction in massive clusters 
\citep[\cf][]{djf95,mohr1999a}.  One can attempt to ameliorate the effects
of departures from this simple picture on the accuracy of the results by determining cosmological
parameters from samples of clusters, in which some of these systematic effects
will average out, and taking care to choose the sample in an unbiased way.  Indeed, much of the observational cluster work cited above (e.g., \citealt{mason2001,grego2001,reese2002}) uses such an approach.  In this work, we investigate the validity of the hydrostatic, isothermal, spherically symmetric assumption by studying observations of a massive (and therefore relatively bright) distant cluster.

We analyze a deep observation of
the galaxy cluster MS1137.5+6625 taken with the \chandra\
Observatory's Advanced CCD Imaging Spectrometer (ACIS).  
\ms1137\ was observed as part of L. VanSpeybroeck's Cycle 1 Guaranteed 
Time
Observations (GTO). These GTO observations, together with observations 
in subsequent \chandra\ observation cycles, form a collection of
X-ray observations of galaxy clusters designed for measuring
cosmological parameters to high accuracy.   
Over 1,250 ks of Chandra time were
scheduled for this project in observation Cycles 1 and 2, and it included over 40 clusters by the close of Cycle 2.
The project was designed to include a sufficient 
number of clusters to overcome systematic errors in the calculated
cosmological parameters arising from cluster ellipticity and
orientation \citep[\cf][]{sulkanen1999}.
The clusters in the sample are massive (kT$_e \gtrsim$ 5 keV) and 
distributed widely in redshift.  The clusters were chosen on the basis
of having high X-ray luminosity and/or high X-ray temperature.  Those
with extremely bright radio point sources at their centers were avoided.  The \ms1137\ observation is one of
the longest observations scheduled for this project.

The Chandra Observatory carries instruments particularly well-suited to studying the 
distant, massive clusters in
the cosmology project, owing to a combination of
unprecedented angular resolution in the X-ray band; low background 
rates, particularly in
the front-illuminated ACIS-I devices; and the ability to detect X-ray
photons up to high energies (kT$_e \sim$10 keV).  The \xmm\ X-ray 
observing facility has substantially more collecting area, permitting investigation of large-scale
temperature gradients in distant clusters with observations of
moderate length.  The \chandra\ facility, however, has significantly
better angular resolution than \xmm, and as is discussed
in this paper, this becomes quite important for distant clusters; their compactness does require the
higher angular resolution of \chandra\ to accurately model the
cluster's spectrum and spatial structure.  The two facilities are
therefore complementary for study of high-redshift clusters.

The \chandra\
observation reveals that while the X-ray image suggests this massive,
distant cluster is relaxed, symmetric, and spherical, its story is 
actually considerably more complicated.
We analyze the \chandra\ data in conjunction with SZE data and compare
it with optical data.  We determine it has a remarkably compact density
distribution, find evidence of asymmetric temperature structure, and measure 
an angular diameter distance which implies a very low value for the
Hubble constant.  

In this paper, we review previous observations of \ms1137\ in
Section~\ref{sec:prevobs} and describe the \chandra\ observations and 
the ICM temperature and density profiles we derive from them in 
Section~\ref{sec:chandraobs}.  In Section~\ref{sec:calculations}, we
derive the ICM cooling time, ICM mass and total mass profiles from the
results of Section~\ref{sec:chandraobs}, and calculate
the cluster distance and infer the Hubble constant from the
X-ray and SZE data in Section~\ref{sec:dist}.  We discuss these results, evaluate the cluster's
suitability for cosmological work, and suggest possible interpretations in Section~\ref{sec:discussion}.

\section{Previous Observations}
\label{sec:prevobs}
MS1137.5+6625 was serendipitously discovered in the \ein\ Extended
Medium Sensitivity Survey (EMSS) \citep{gioia1990} and was
subsequently identified
as a distant galaxy cluster \citep{henry1992,gioia1994}.  It was 
confirmed
by \cite{donahue1999} to be at a redshift of z=0.784.  
Assuming \omegam=0.3, \omegal=0.7, $h$=0.65 (which we will use 
throughout this paper), this
redshift implies an angular diameter distance (\da) of 1655.50
$h_{65}^{-1}$ Mpc and a scale of 8.0 kpc arcsec$^{-1}$.

Further observations of the cluster confirmed that it is massive and 
compact.
From a 70 ks \asca\ observation, \cite{donahue1999} find the
cluster's best-fit rest-frame emission-weighted intracluster medium
temperature to be kT$_e =$ 5.7$^{+2.1}_{-1.1}$ keV, at 90\% confidence.
From Keck II spectra of 22 cluster members, they ascertain a velocity 
dispersion of 884$^{+185}_{-124}$ km s$^{-1}$.
\cite{clowe2000} obtained deep optical {\it R}-band images of 
\ms1137\
with the Keck II 10 m telescope and {\it I}-band images with the
University of Hawaii 2.2 m telescope.  The optical light of the
cluster has a compact distribution, and the centroid of this 
distribution is
coincident with the brightest cluster galaxy (BCG).  There are two 
giant strong gravitational lensing arcs, at 5.5$''$ and 18.0$''$ from the BCG, and
several other arc candidates, indicating a high surface mass density.
\cite{clowe2000} performed a weak lensing analysis of the data, and 
found
the mass distribution is as compact as the light distribution.  Using
aperture densitometry, they found the minimum cluster mass within
a radius of 2$'$ centered on the BCG to be $(4.9\pm1.6)\times
10^{14} h_{65}^{-1} $M$_\odot$.  A second mass peak to the north of the 
cluster may
or may not be associated with the cluster, but only contributes about
10\% of the total mass.

\section{\chandra\ Observations}
\label{sec:chandraobs}
\begin{figure}
\plotone{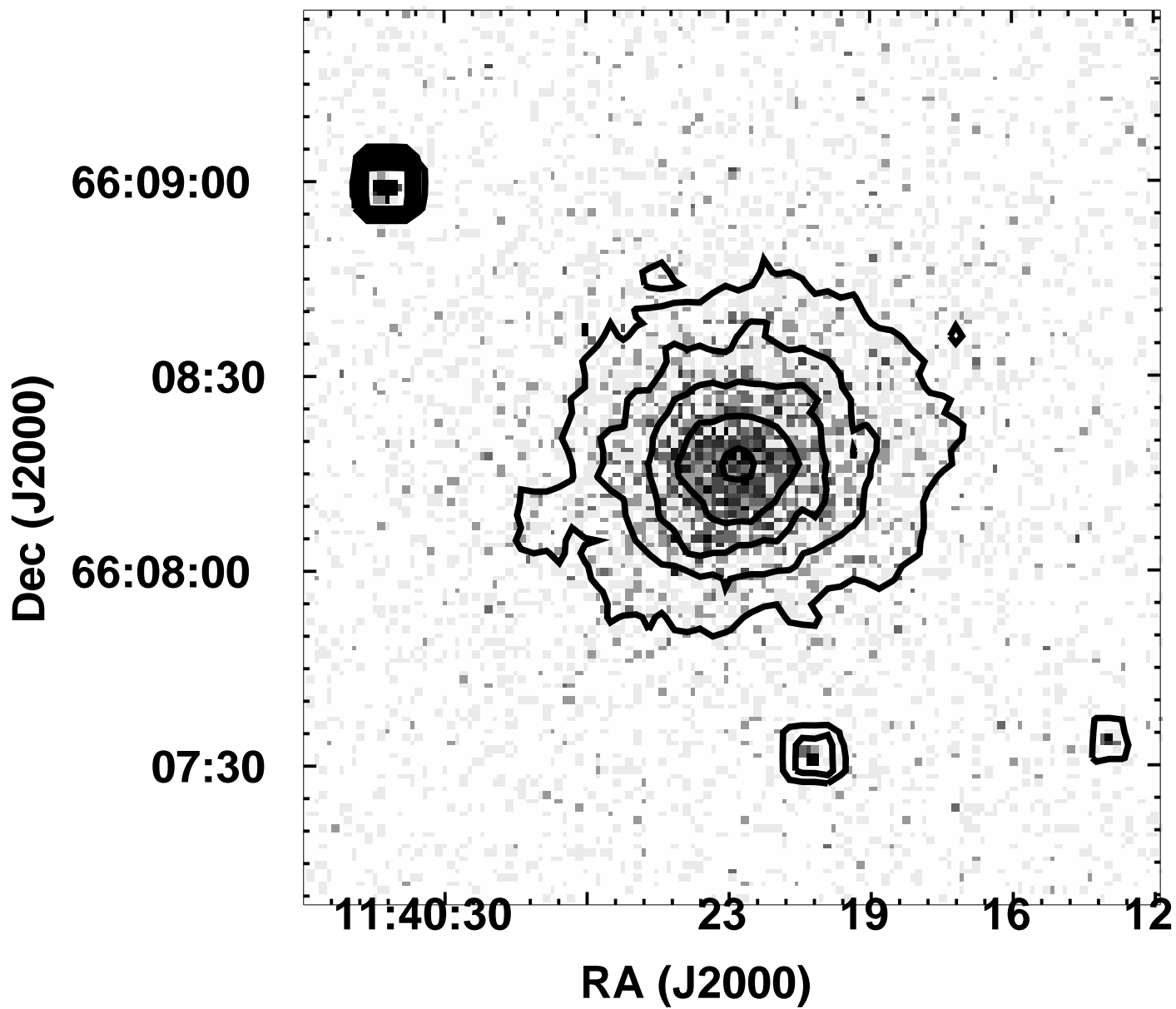}
%\plotone{ms1137_bin4_2.cps}{clusterimagewithcontours.ps}
\caption{Raw 0.3 keV to 10.0 keV image of MS1137.5+6625, blocked into 2$''$ pixels.  There are approximately 4,000
cluster counts in 107 ks of time.  No background subtraction has been
performed. The contours are logarithmically spaced from 0.55 to 30 
counts, where the map has been smoothed over 4 pixels.  The aimpoint is 
$\sim$ 1 arcmin south and $\sim$2.5
arcmin east of the cluster's center; the instrument PSF is nearly
unchanged from the aimpoint at the cluster's position.}
\label{fig:rawimage}
\end{figure}

\ms1137\ was observed from September 30 to October 2, 1999 for an 
elapsed
 live time of 117.712 ks.   The aimpoint of the ACIS-I3 chip was $\sim 
1'$ south and 2.5$'$
 east of the cluster center.  We selected events with the standard
 \asca\ grades and rejected data taken during times with background
 rates more than 20\% greater than the mean.  The effective exposure 
time after this filtering was 107.045 ks.
 In Figure~\ref{fig:rawimage}, we present a raw image, without a
background subtracted,  in the 0.3 to
 10.0 keV energy range, blocked into 2$''$ pixels.  There are about
 4000 cluster photons in the image. The background rate in the map is 
about 0.15
 counts arcsec$^{-2}$; the average emission rate over the cluster area is about four times 
brighter.
In the image, the cluster looks relaxed and symmetric.
\subsection{Temperature Structure}
\label{subsec:temperature}
To determine the cluster's emission-weighted temperature, we extract
the cluster spectrum within 63$''$ (506 $h_{65}^{-1}$ kpc) from the 
cluster's center. We estimate the
background in the same region using a number of long integrations on
blank fields compiled by Markevitch.  (See 
http://hea-www.harvard.edu/$\sim$maxim/axaf/acisbg/ for a complete
description of the background datasets and their reduction.)  We use 
the background data set taken coevally with the
\ms1137\ observations, so that the \chandra\ focal plane was at the 
same temperature ($-$110 C).  
We use Vikhlinin's {\it calcrmf} package to derive response
matrices and effective area files for the cluster spectrum; this 
software ensures that the
variation of the energy response across the I3 chip is taken into 
account.
Before extracting the spectra, we mask out the pointlike sources in
the observed field.  This has little effect on the number of
cluster photons we can retrieve;  only one source, $\sim$45$''$ southwest
of the cluster center, was within the cluster spectrum extraction region.
The spectrum contains $\sim$4200 cluster photons and $\sim$1800 
background photons.

We fit this spectrum, grouped into channels with a minimum
of 20 counts to ensure applicability of the \chisq\ fit statistic, to
a single-temperature MEKAL plasma model in the 0.7-7.5 keV energy
range, using XSPEC.  The lower energy bound is chosen so that we keep
only the best-calibrated data.  We make our best estimate of the
background spectrum by an iterative process of varying the
background spectrum normalization and
noting how this affects the \chisq\ fit statistic for the best-fit
model.  Since the hard and soft components of the X-ray background
originate in different physical processes (cosmic X-rays
dominate the background at energies below $\sim$5 keV and cosmic rays dominate above
$\sim$5 keV), the background in the \ms1137\ observation may depart
from the blank-sky background differently at low and high energies.
To mitigate these effects while keeping the high-energy cluster photons,
we choose
an upper bound of 7.5 keV for our fit.  We find the best estimate of the background
to be 4.5\% higher than the blank-sky background file for our epoch.

\begin{figure}
\centerline{\hbox{
\psfig{figure=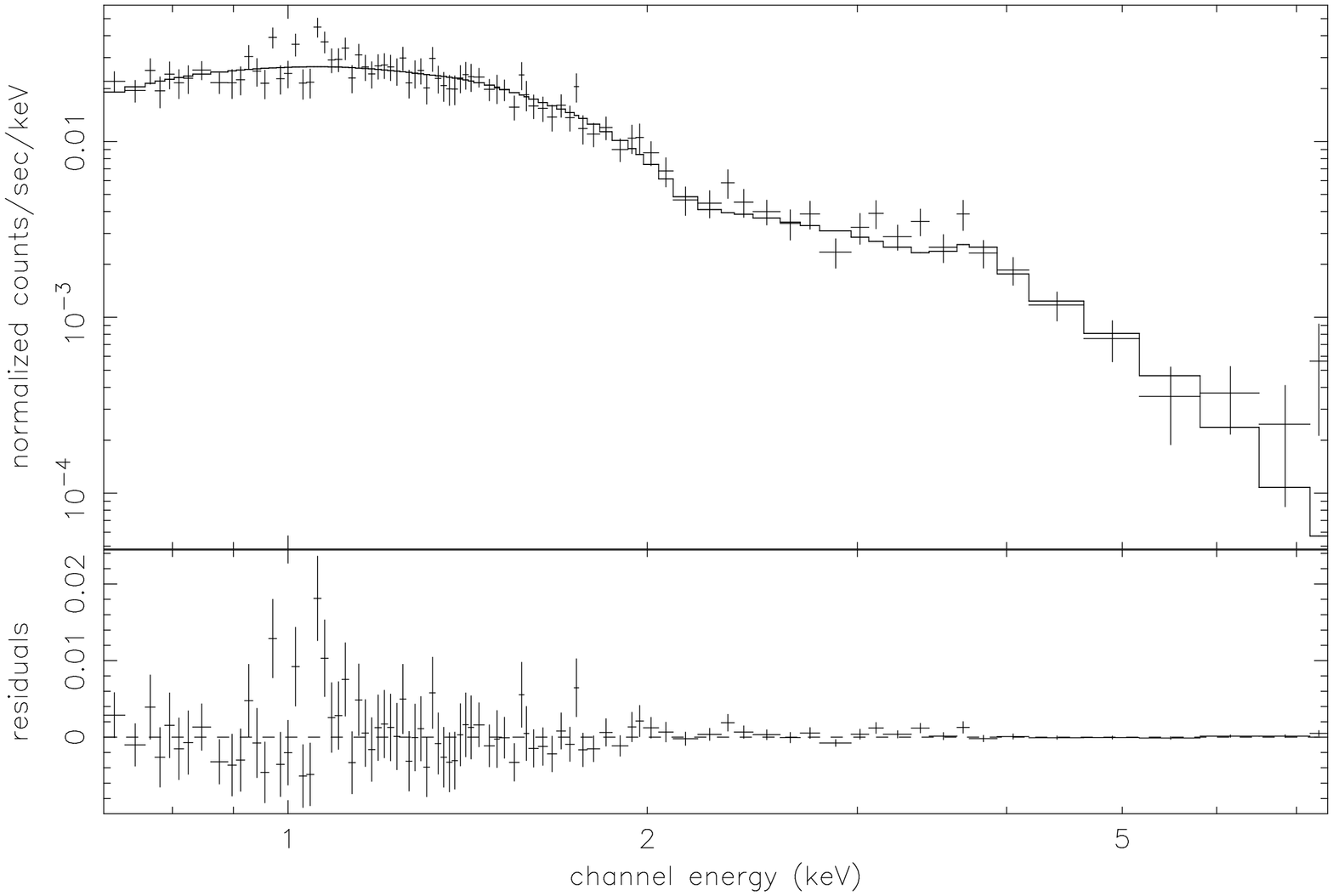,height=2.2in,angle=270}
\psfig{figure=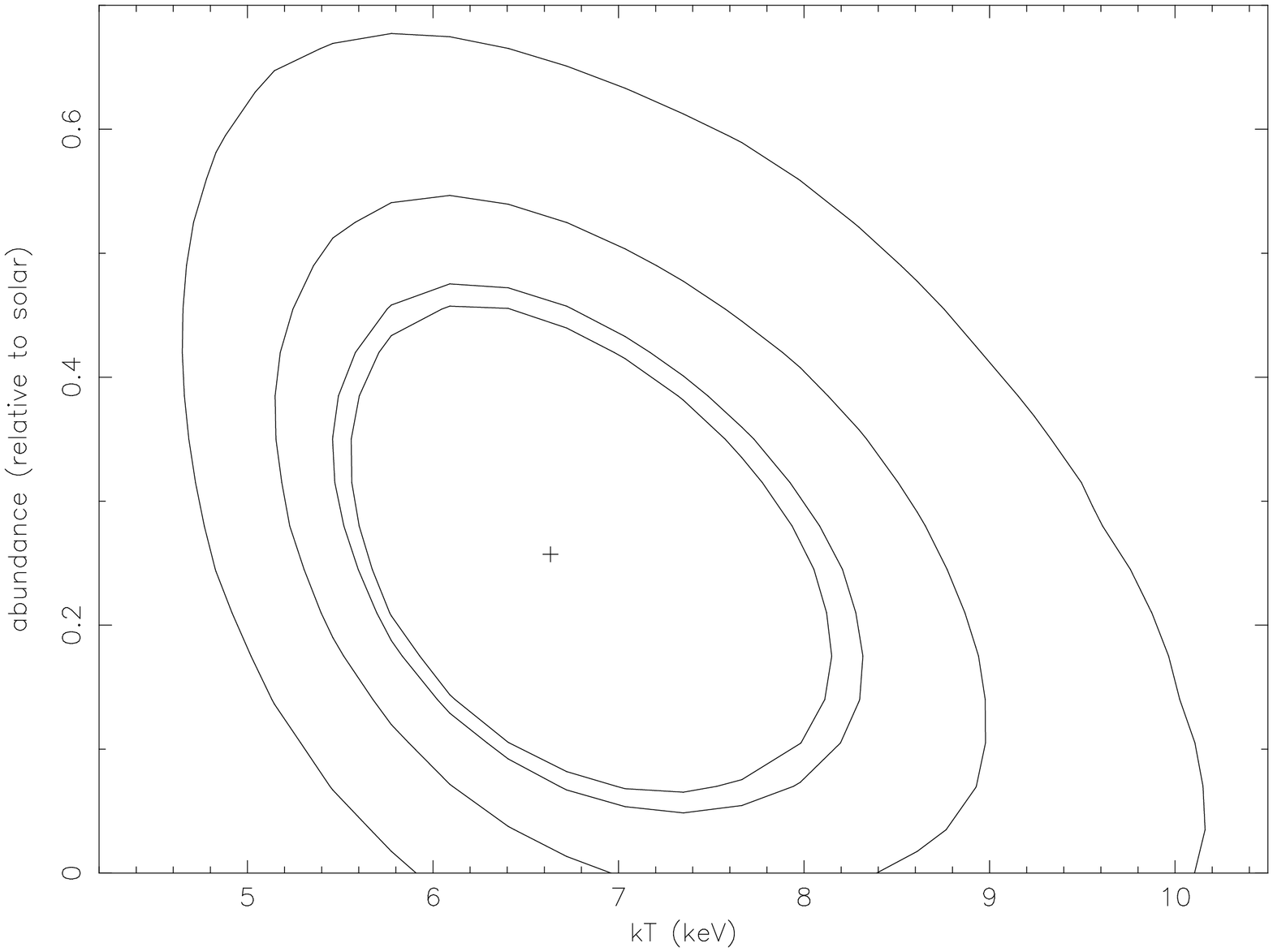,height=2.2in,angle=270}
}}
%\plottwo{ms1137_07to75fit.ps}{kT_vs_abund_outer_07to75.ps}
\caption{Left panel: the \chandra\ ACIS-I spectrum, best-fit model,
and residuals for the full cluster spectrum for \ms1137.  Right panel:
the two-parameter confidence region for emission-weighted temperature
(kT$_e$) and abundance. The first, third, and fourth regions
correspond to 68.3\%, 90\%, and 99.0\% confidence on the two
parameters jointly.  The projection of the second region onto either axis
marks the 90\% confidence interval on the single parameter.  The `+' 
marks the best-fit value.}
\label{fig:kTvsabund}
\end{figure}

We find the best-fit emission-weighted mean temperature for the
intra-cluster medium to be 6.64$^{+1.67}_{-1.20}$ at 90\%
confidence.  The best-fit metal abundance is 0.26$^{+0.21}_{-0.21}$
times the solar abundance as determined by \cite{anders1989}, again at
90\% confidence.  The cluster spectrum, its best-fit model, and the
fit residuals are shown in Figure~\ref{fig:kTvsabund} along with the two-parameter confidence regions for
 temperature and metal abundance.  The neutral hydrogen column density is
not strongly constrained by the data in this energy range. The 90\% 
confidence interval is $n_H = 
2.28^{+3.76}_{-2.28}\times 10^{20}$cm$^{-2}$, consistent with the
Galactic value for this direction: 1.14$\times
10^{20}$cm$^{-2}$ \citep{dickey1990}. The reduced \chisq\
statistic for this fit is 1.051 for 150 degrees of freedom.

Since the density profile suggests that the gas in the cluster center may
be dense enough to cool (as discussed in Section~\ref{sec:calculations}), we 
also fit the
cluster's spectrum to a MEKAL plasma plus a MEKAL cooling flow 
component.  We fix the lower cutoff temperature to 0.2 keV and tie the
upper temperature of the cooling flow to the MEKAL plasma temperature.  The best fit
cooling flow upper temperature
is consistent with the emission-weighted temperature we found above,
the cooling flow mass deposition rate is consistent with zero, and the
reduced \chisq\ is not improved.  We conclude that there is no
spectral evidence for a cooling flow component in the cluster.

We compare these results to those obtained by
\cite{donahue1998,donahue1999}, who found kT$_e$ = $5.7^{+2.1}_{-1.1}$ 
keV
and an iron abundance of $0.43^{+0.40}_{-0.37}$ times solar abundance,
at 90\% confidence.  These results are
consistent with the \chandra\ results.  Some difference may be 
expected,
as there are a number of bright point sources in
this field which can be distinguished by \chandra\ and their
emission removed
from the cluster spectrum, but which would likely contaminate the
 \asca\ cluster spectrum.  The point sources in
the I3 chip have a background-subtracted sum of about 3800 counts in
the 0.7-10.0 keV band. The brightest of these pointlike sources,
located $\sim$ 60$''$ to the northeast of the cluster center,
contributes $\sim$ 1000 of the 3800 counts.  This emphasizes the 
importance of good angular
resolution for studies of high-redshift clusters; it is difficult
to determine accurately the ICM temperature and luminosity of high-redshift clusters
without the ability to exclude point sources.

\begin{figure}
\plotone{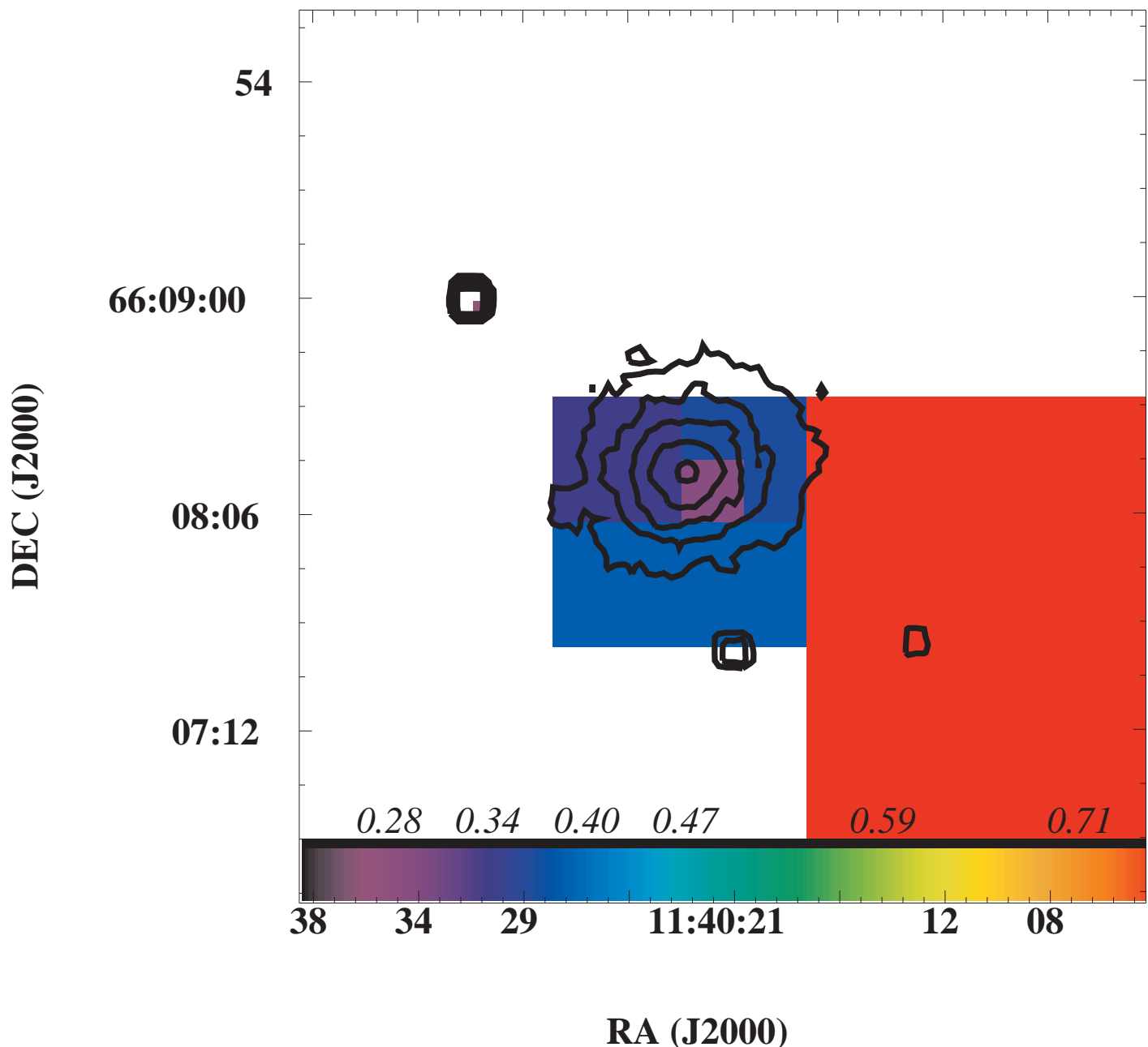}
\caption{Hardness-ratio map of MS1137.5+6625.  The cluster image in
the 2.5-5.0 keV energy range is divided by the 0.5-2.5 keV image.  The
background is subtracted from each image and the fractional error is a 
constant value
of 0.15 across the map.  An emission-weighted temperature and
abundance of 7.0 keV and 0.245 solar would give a
hardness ratio of $\sim$ 0.165.  Lower values correspond to lower
temperatures and/or metallicity.  The surface-brightness contours from
Figure~\ref{fig:rawimage} are overlaid.}
\label{fig:hardness}
\end{figure}

To investigate the presence of structure in the temperature or metallicity
distribution in this cluster, we make a hardness ratio map using the 
adaptive binning technique described in
\cite{sandersfabian2001}.  This technique ensures a constant
signal-to-noise ratio across the map;  we set the fractional error in
each bin to be 0.10. We estimate the background in each band from the
\ms1137\ dataset.  In 
Figure~\ref{fig:hardness} we show a map of the cluster in the 2.0-5.0
keV band divided by the 0.43-2.0 keV band with the surface brightness
contours of Figure~\ref{fig:rawimage} overlaid.  This ratio is 
sensitive to
temperature variations, moderately sensitive to metal abundance
variations, and essentially insensitive to neutral hydrogen absorption
variations. 
There appear to be significant fluctuations, with the center of the
cluster softer than the outer regions.  The hardness ratio image appears more 
complicated than the raw image, suggesting variations
in temperature and/or abundance may not be present in the ICM 
density.  Such behavior is not uncommon in clusters at low redshifts.  
\cite{donnelly2002} derive temperature maps for a
complete sample of nearby clusters observed with the \asca\ facility.
Many of these clusters which appear regular in surface
brightness, even in images with much better photon statistics, exhibit quite complicated temperature structure,
suggesting that temperature and abundance variations, likely
created during cluster merging, are either created frequently or 
persist for a long time.

We can quantify the structure in this map further by identifying the
hardness ratios with those expected from a MEKAL plasma.  A MEKAL 
plasma at our best-fit emission-weighted
temperature of 6.7 keV and metallicity of 0.25 solar gives a hardness
ratio of $\sim$0.28 in these bands.  For a hardness ratio of 0.35, a
$\sim$ 7 keV plasma must have an abundance of $\sim$ 1.0; or, if the
abundance is to be 0.25, the plasma must be at a temperature of
$\sim$10 keV.

We attempt to confirm this temperature variation by fitting spectra
extracted from different regions.  The first is the region within 17$''$ and the second is the annulus
between 17$''$ and 47.5$''$.  The best fit temperature and abundance
for the inner region are 6.9$^{+2.2}_{-1.5}$ keV and
0.21$^{+0.26}_{-0.21}$ times solar, at 90\% confidence.  The best fit temperature and abundance
for the outer region are 8.8$^{+3.5}_{-2.6}$ keV and
0.28$^{+0.39}_{-0.28}$ times solar.  The data are consistent with a cooler inner region, but this is not statistically significant at 90\% confidence.
There are not enough photons to support a temperature deprojection
analysis.
%We attempt to confirm this temperature variation by fitting spectra
%extracted from different regions.  We divide the region within 63$''$
%in half along a line rotated $\sim$45$^\circ$
%counterclockwise from the east-west line through the cluster center.
%The best-fit MEKAL model to the northern half gives temperature
%7.97$^{+4.13}_{-2.38}$ keV and abundance 0.12$^{+0.37}_{-0.12}$ solar;
%the southern half gives a temperature 6.40$^{+1.10}_{-1.5}$ and
%abundance 0.05$^{+0.25}_{-0.05}$ solar.  Although the two regions are
%formally consistent with each other at the 90\% confidence level, the
%spectral analysis does bear out the implications of the hardness ratio
%map: the northern region is probably hotter with higher metallicity than the
%southern region.  

\subsection{Density Profile}
\label{subsec:density}
We parameterize the spatial distribution of the electron number density 
$n_e$ as a
spherically-symmetric isothermal
beta-model\citep[\cf][]{cavaliere1976,cavaliere1978}.  The beta-model
has the form:
\begin{equation}
n_e(r) = n_{e0} \left ( 1 + \frac{r^2}{r_c^2} \right )^{-3\beta/2},
\label{eq:iso_beta}
\end{equation}
where $n_{e0}$ is the electron number density at the cluster's center;
\be, though derived with a physical meaning, is in practice a fitted
parameter; $r_c$ is a characteristic length scale for the cluster,
also a fitted parameter; and $r$ is the radius from the
center. Thermal bremsstrahlung X-ray emission from such
a plasma results in a surface brightness profile
of the form
\begin{equation}S_X(\theta) = S_{X0} \left (1 + 
\frac{\theta^2}{\theta_c^2} \right )^{\frac{1}{2}-3\beta}\label{eq:betamodel},\end{equation}
where $S_{X0}$ is the surface brightness at the center of the cluster,
$\theta_c$ is the cluster's projected core radius
$\theta_c= {{\rm r}_c/D_A}$, and $\theta$ is the angular distance from
the cluster center.  Using the CXC Sherpa package, we fit an image made 
in the energy range
0.3 keV to 5.0 keV to a generalized beta-model, in
which two perpendicular axes can have different core radii.  (As noted
by \cite{rybicki1984}, a biaxial ellipsoid with a beta-model
electron-density distribution will project to a simple
elliptical beta-model of this type for any orientation of the axes with respect to
the observer.)  The axis ratio, i.e., the ratio of the minor to the
major axis, is 0.87$^{+0.06}_{-0.07}$, at 90\% confidence.
  We use the image
analysis to quantify the deviation from circular symmetry and to locate 
the cluster's
center. However, using a biaxial or triaxial cluster model
introduces a considerable amount of ambiguity, \cf\
\cite{grego2000a}.  We use a spherically symmetric model, checking
that the best-fit shape parameters from the two-dimensional fit are
consistent with those from the spherically symmetric model, and leave the discussion of the effects of deviations 
from spherical symmetry until Section~\ref{sec:discussion}.	

\begin{figure}
\plottwo{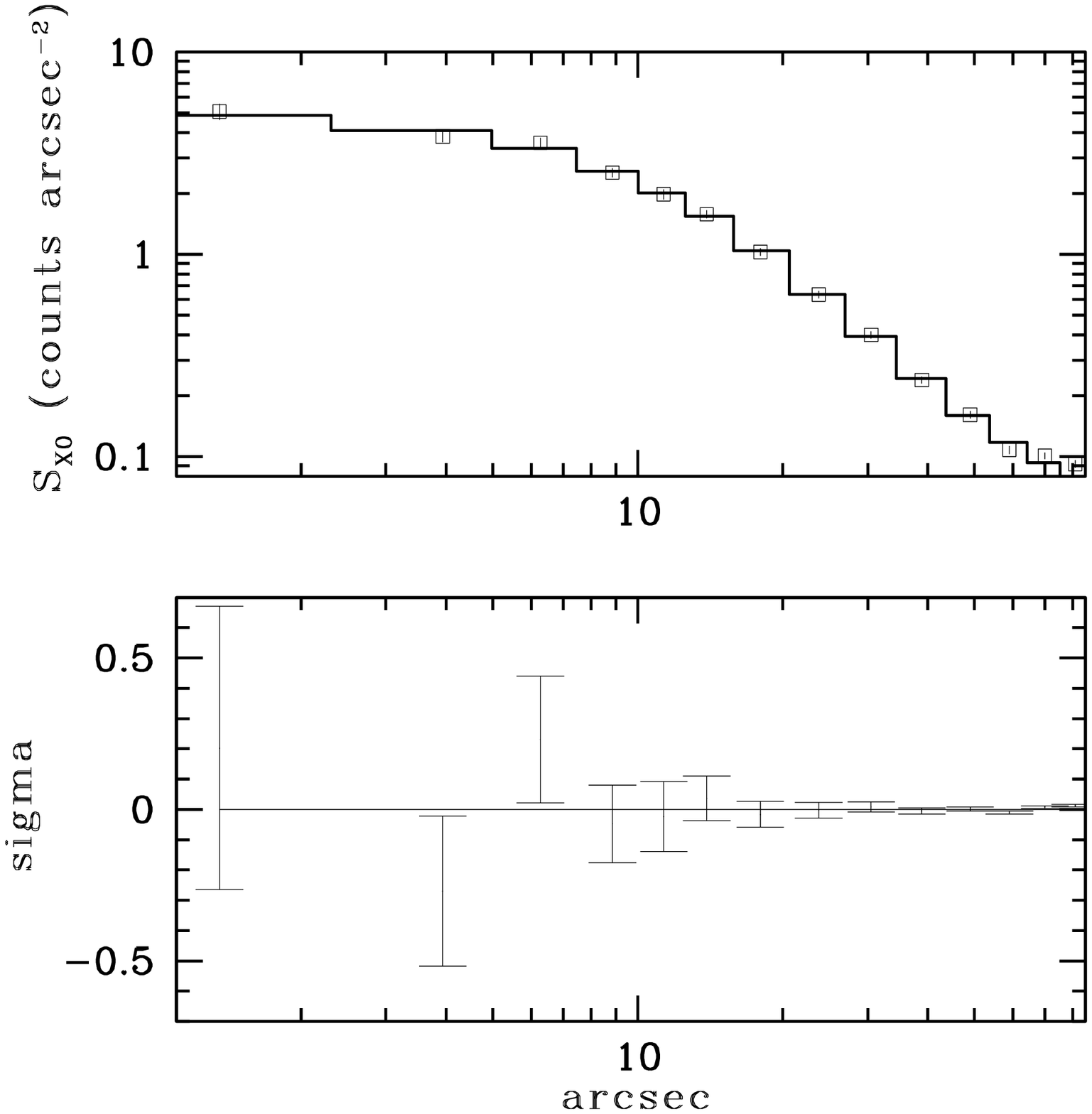}{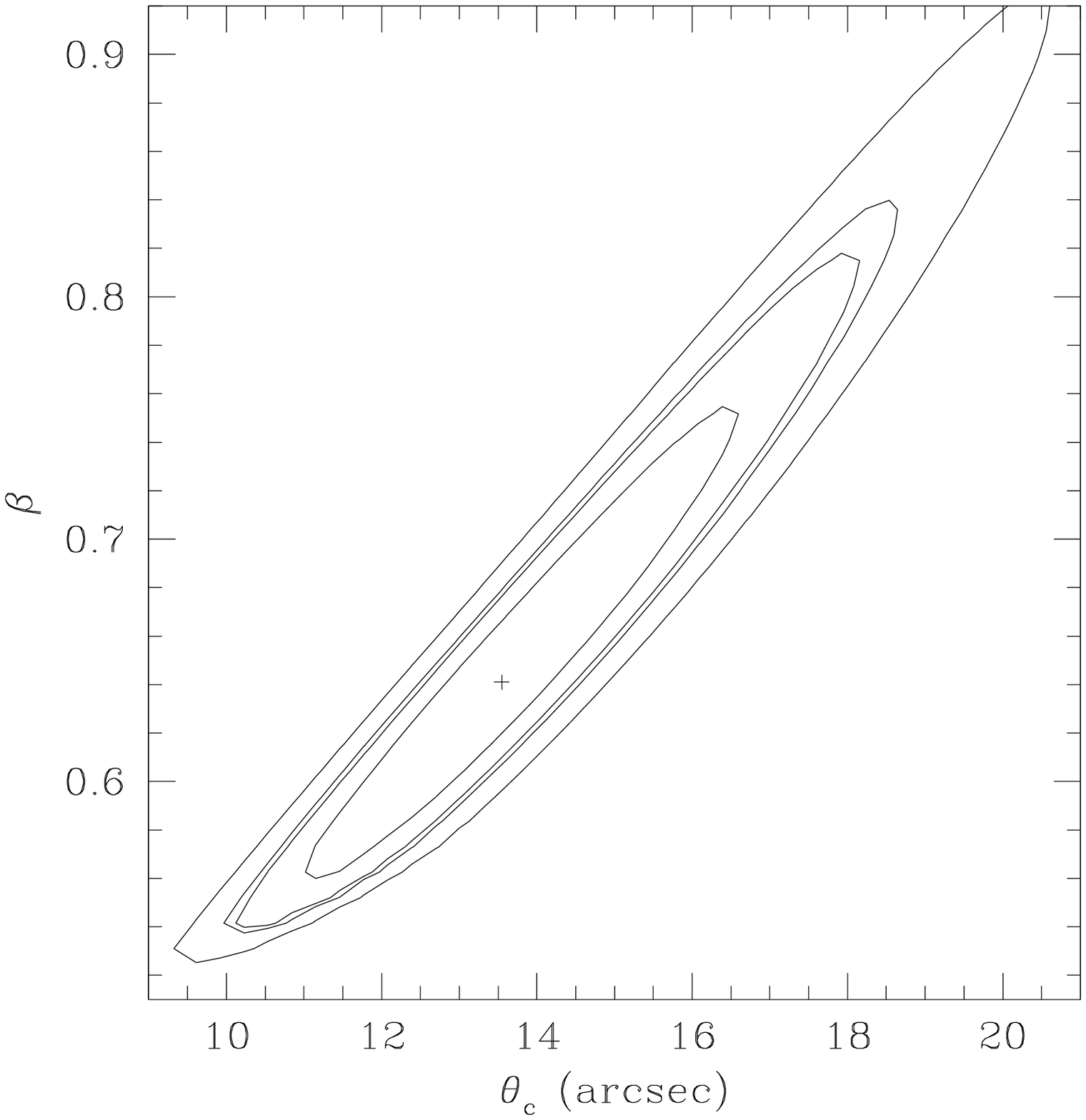}
%\plottwo{one_d_fit_july9.ps}{confidence_beta_rc_july9.ps}
\caption{Left panel: One-dimensional surface brightness profile of 
cluster MS1137.5+6625 with background, its best-fit beta-model, and fit 
residuals in units of the standard deviation $\sigma$..  Right
panel: Two-parameter confidence region for $\beta$ and $\theta_c$
for the one-dimensional beta-model fit to MS1137.5+6625.
The confidence contWe use values \omegam\ = 0.3
and \omegal = 0.7.ours correspond to 68.3\%, 90\%, 93\% and 99.0\%
confidence on the two parameters.  The projection of the 93\%
confidence interval onto an axis gives the single-parameter 90\%
confidence interval.  The best-fit model is marked with a `+'.}
\label{fig:betarcfit}
\end{figure}

Using the centroid from the two-dimensional fit (RA =
11$^h$40$^m$22$''$.4, Dec.= +66$^\circ$ 08$'$ 14$''$.7), we
derive a one-dimensional surface brightness profile and
fit it to the beta-model of Equation~\ref{eq:betamodel}.  We model the
background (as a constant) concurrently with the cluster model, so
that we can use the log-likelihood statistic (called the ``Cash''
statistic in Sherpa).  This statistic,
$S \equiv -2 ln(L)$ where $L$
is the Poisson likelihood, tends asymptotically to be distributed like the \chisq\ statistic, so
confidence intervals on the calculated quantity can be
constructed by the deviation from the minimum statistic.  Over the small 
angular region within which we are fitting
(within a radius of 85$''$) and in this energy band, the effective
area of \chandra\ changes by less than 3\% and the background (as estimated from the 
Markevitch
background files) is constant.   We also allow for
an offset between the centroid we determined from the image analysis and the surface brightness 
peak.  In Figure~\ref{fig:betarcfit} we show the data, the best fit model, 
and the
fit residuals; the best-fit model has $\beta$=0.63 and a core radius
of 13.3$''$, or 106 $h_{65}^{-1}$ kpc.  The 
confidence intervals on the surface-brightness shape parameters
 are shown in the right panel of the figure.  For the confidence
interval calculation, the offset between the centroid and peak is
fixed to its best-fit value of $2.1''$, reasonable for a cluster with the observed
axis ratio.  If we require the peak to coincide with the centroid, the 
best-fit core-radius decreases to 10.9$''$, the
best-fit $\beta$ is 0.56, and the fit statistic degrades by $\Delta 
S$=1.6.  The small core
radius of this cluster reinforces the importance of Chandra's high
angular resoution for studying high-redshift clusters.

We fit this profile again, after eliminating the central region, to see 
if the
small (and marginally statistically significant) central excess is
affecting the shape parameter fit, and we find no appreciable change
in the fitted parameters.

We recover the central gas density from the normalization of
the fitted XSPEC model to the spectrum.
The XSPEC normalization is as follows:
\begin{equation}N_{XS} = {10^{-14}\over 4\pi(1+z)^2 D_A^2}\int n_H 
n_e dV,\label{eq:xspecnorm}\end{equation}
where the integral is taken over the volume projected onto the area
within which the spectrum was extracted, here, within a circular
region with radius R. All variables are in cgs units.  The central density 
can be found from this relation: 

\begin{eqnarray}
n_{e0}^2& =&{4\, \pi \, (1+z)^2 D_A\, ^2 N_{XS}\over 10^{-14}\, (n_H/n_e)}
\left({1-3\beta\over 2\, \pi r_c^3 }\right)\nonumber\\
&\times& \left[ \int\limits_0^\infty \left\{ \left(1+\zeta^2+ {R^2 \over
r_c^2}\right)^{1-3\beta}\hspace{-5pt} - \left ( 1+\zeta^2 \right
)^{1-3\beta}\right\} d\zeta \right]^{-1},\label{eq:ne0}\end{eqnarray}
where $\zeta$ is the distance from the cluster
center along the line of sight in units of the core radius.

To determine the uncertainty in the central electron density 
measurement due to the spatial
fits, we calculate n$_{e0}$ at a large number of parameter pairs (grid
points)
distributed over the
90\% confidence interval on \rc, \be\ of Figure~\ref{fig:betarcfit}b.
Determining the confidence interval on $n_{e0}$ from the relevant
confidence area  for \rc\ and \be\ is necessary to correctly account 
for the strong correlation
between parameters.   We add in quadrature the spectrum normalization
uncertainty and the uncertainty from the spatial fits.  We find a 
central density of
0.0136$^{+0.0016}_{-0.0013}$ cm$^{-3}$.  In Figure~\ref{fig:nefgvsr}, 
we illustrate the relative contributions of the uncertainties; we plot
with a solid line the $n_e(r)$ calculated for each grid point within
the 90\% confidence interval (the contribution from the spatial fit) and with
a dashed line the 90\% confidence
interval on $n_e(r)$ which includes the contribution from the 
uncertainty in the XSPEC normalization.  We note that uncertainty is
dominated by the uncertainty in the shape parameters.

\section{Calculated Quantities}
\label{sec:calculations}
\subsection{Cooling Profile}
\label{subsec:cooling}
\begin{figure}
\plottwo{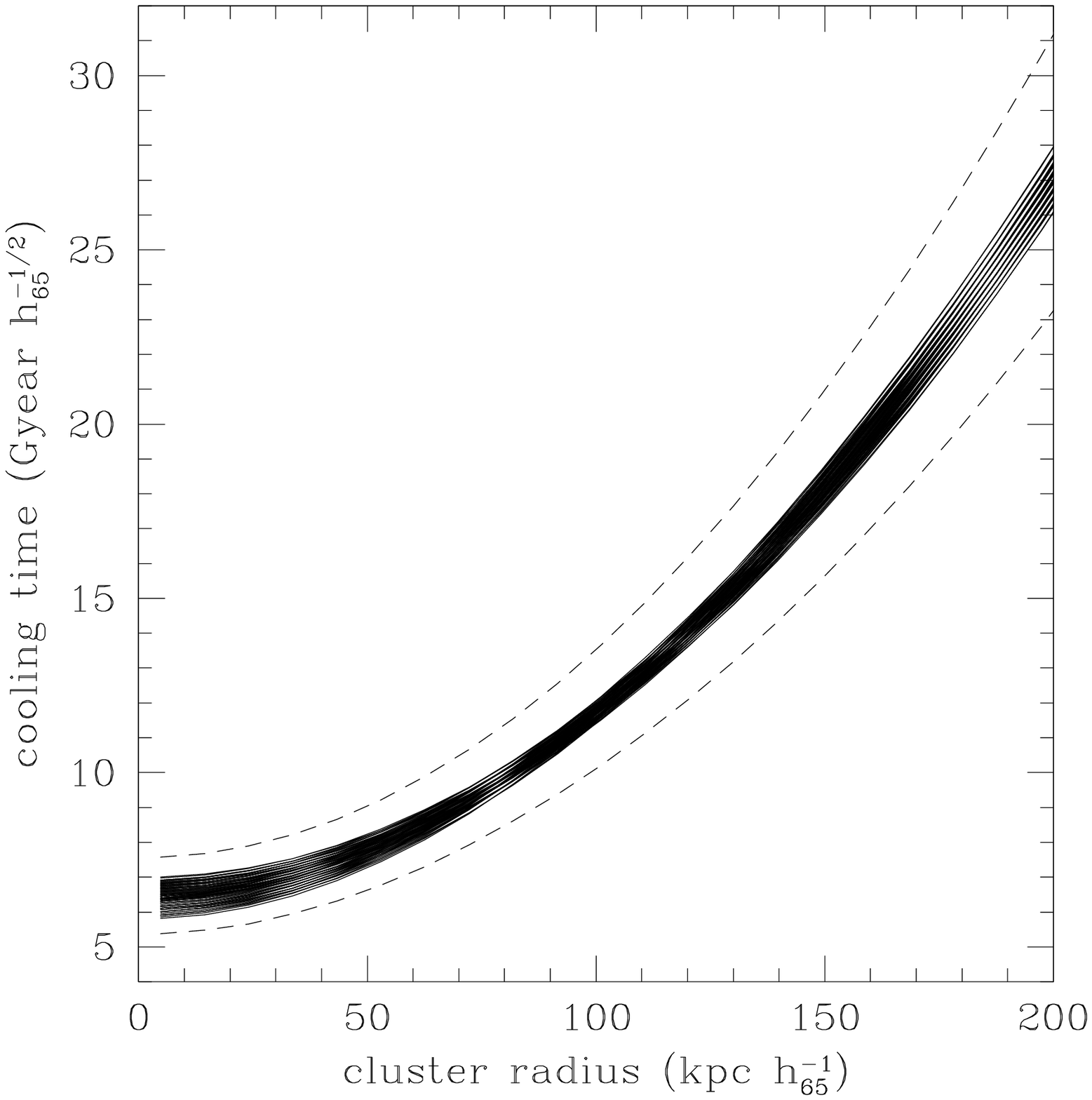}{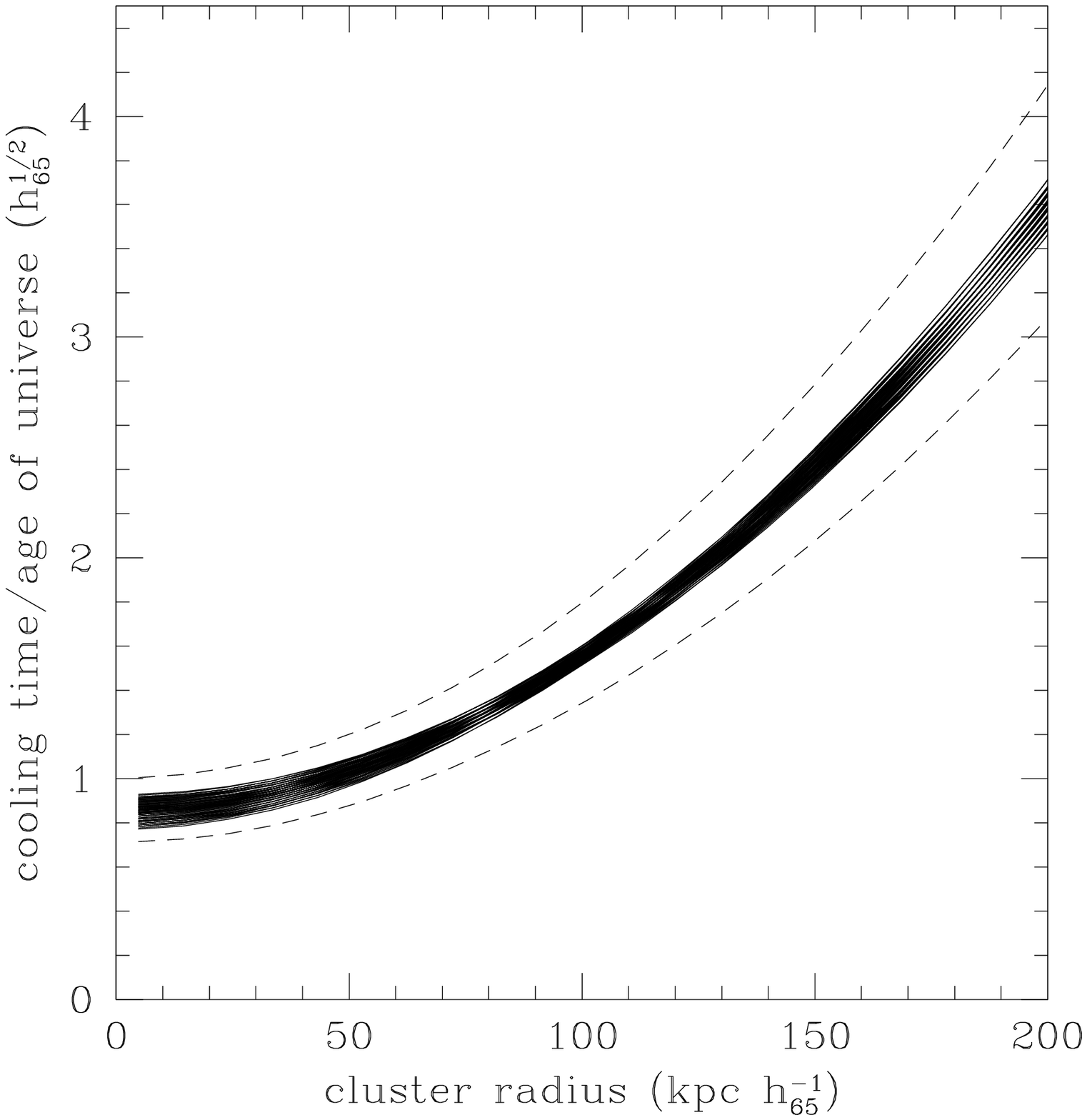}
%\plottwo{cooling_vs_rkpc_65.ps}{cratio_vs_rkpc_h65.ps}
\caption{Left panel: The cooling time of the ICM in Gyr as a function
of cluster radius in kpc. Right panel: The cooling time of the ICM in 
units of the age of the
universe at the cluster's redshift, as a function of cluster
radius in kpc.  The calculations are done using values \omegam\ = 0.3
and \omegal = 0.7.  Again, the solid lines show the cooling times
calculated for the parameter grid
points within the 90\% confidence region of the fitted model and the
dashed lines show the full 90\% confidence intervals including the
uncertainty in $N_{XS}$ and $T_e$.  The age of the universe at the 
cluster's epoch is 7.53 Gyr $h_{65}^{-1}$. }
\label{fig:tcoolvsr}
\end{figure}

We calculate the cooling time of the ICM as a function of
cluster radius, using the ratio
$(5kT_e/2)/(\epsilon_{ff}(T_e)n_e(r))$, where $\epsilon_{ff}(T_e)$ is
the free-free emissivity of a plasma with our best-fit abundance (line
emission contributes very little cooling at 7 keV).  
The cooling time in Gyr as a function of cluster radius is shown in
the left panel of Figure~\ref{fig:tcoolvsr}.  In the right panel, we 
compare the cooling time to the age of the universe at the cluster's
epoch, which is 7.53 $h_{65}^{-1}$ Gyr in our chosen cosmology.   The 
cooling
time of the ICM appears to be shorter than the cluster's age at radii
below 25-70 $h_{65}^{-1}$ kpc.  However, we do not see convincing
evidence for a cooling flow in the cluster's image or spectrum.  
For the cooling time in the center of the cluster to be longer than the
cluster's age, one or more of our assumptions must be incorrect.
There could be a source of heat to the gas at the cluster center
(although there is no evidence for a radio-loud galaxy there).
The cosmology we use could be incorrect, but for the cooling
time to be longer than the cluster's age with certainty, $h$ must be greater than 1.0 for an \omegam=0.3 flat universe, or
\omegam=0.8 if $h$ is $\sim$ 0.65 in a flat universe.
It is also possible that our simplified physical model for the cluster
is incorrect; if the gravitational potential is steeper along the line
of sight than it appears in projection, the central density we
calculate would be overestimated, and the cooling time
underestimated.  A relatively modest change of
\rc(line of sight)/\rc(projected) = 0.8 can lower the central density by
$\sim$50\%.  Or the cluster may have formed or undergone a merger with enough power
to disrupt cooling processes at some time more recent than $t_{cool}$.

\subsection{Gas and Total Mass Profiles}
\label{subsec:masses}
The gas mass profile follows directly from the electron number density
profile \citep[\cf][]{sarazin1988}:
\begin{equation}M_g(R) = 4 \pi \int\limits_0^{R} n_e(r)\, \mu_e m_p\, 
r^2\, dr.\label{eq:mgas}\end{equation}  
We use the best-fit abundance to calculate the mean atomic mass $\mu
m_p$ and assume $\mu$ is constant throughout the gas, and that the
plasma is fully ionized.  
We calculate the
cluster's total mass under the assumption that the ICM is
isothermal and in
hydrostatic equilibrium in the cluster potential, with thermal support
only.  Under such assumptions, the total mass of a cluster within 
radius $R$ is \begin{equation}M(R) = {3 k T_e \beta \over G \mu m_p} 
{R^3\over
r_c^2 + R^2},\label{eq:totalmass}\end{equation}  
where G is the gravitational constant.

\begin{figure}
\epsscale{0.80}
\plottwo{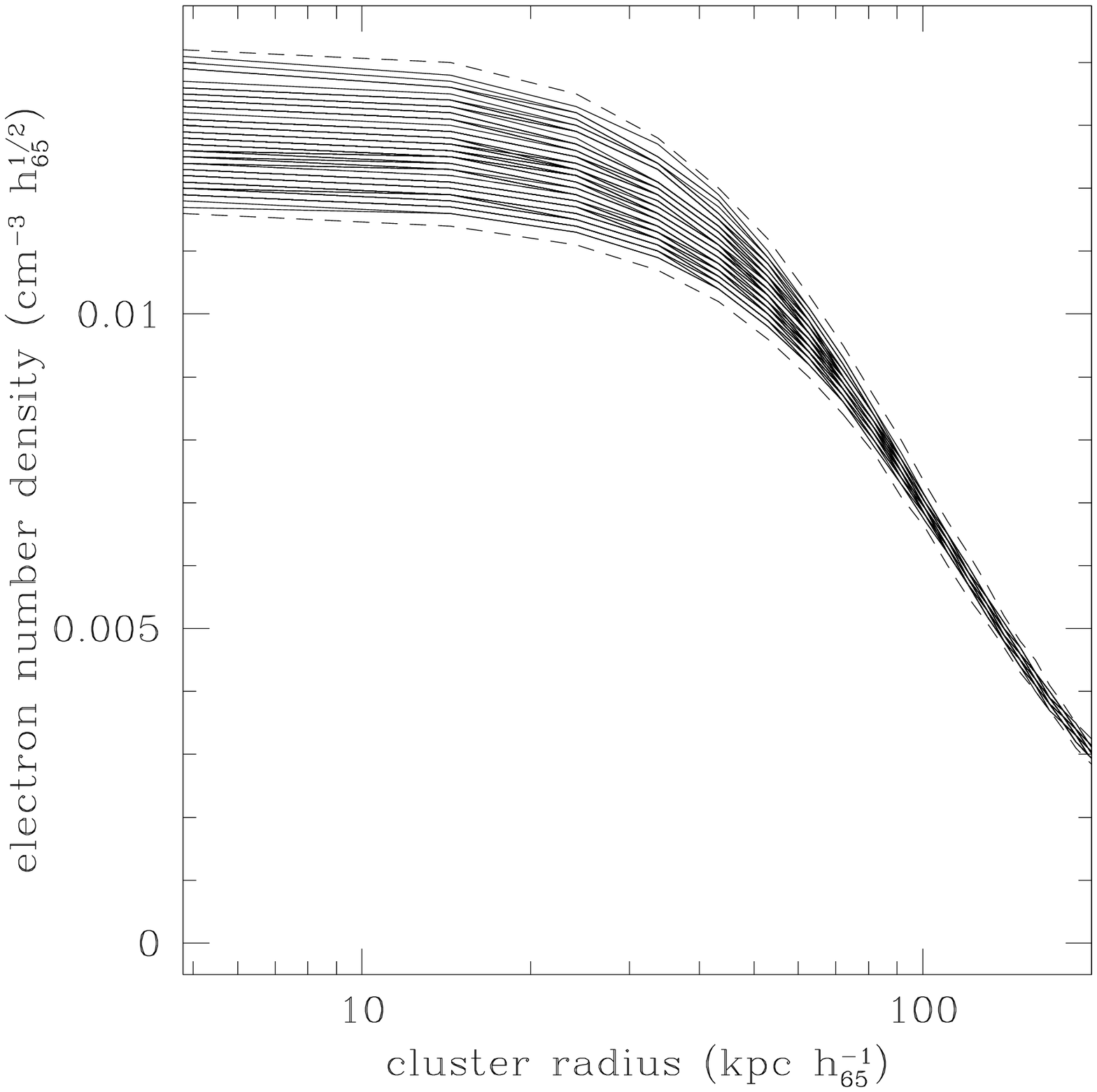}{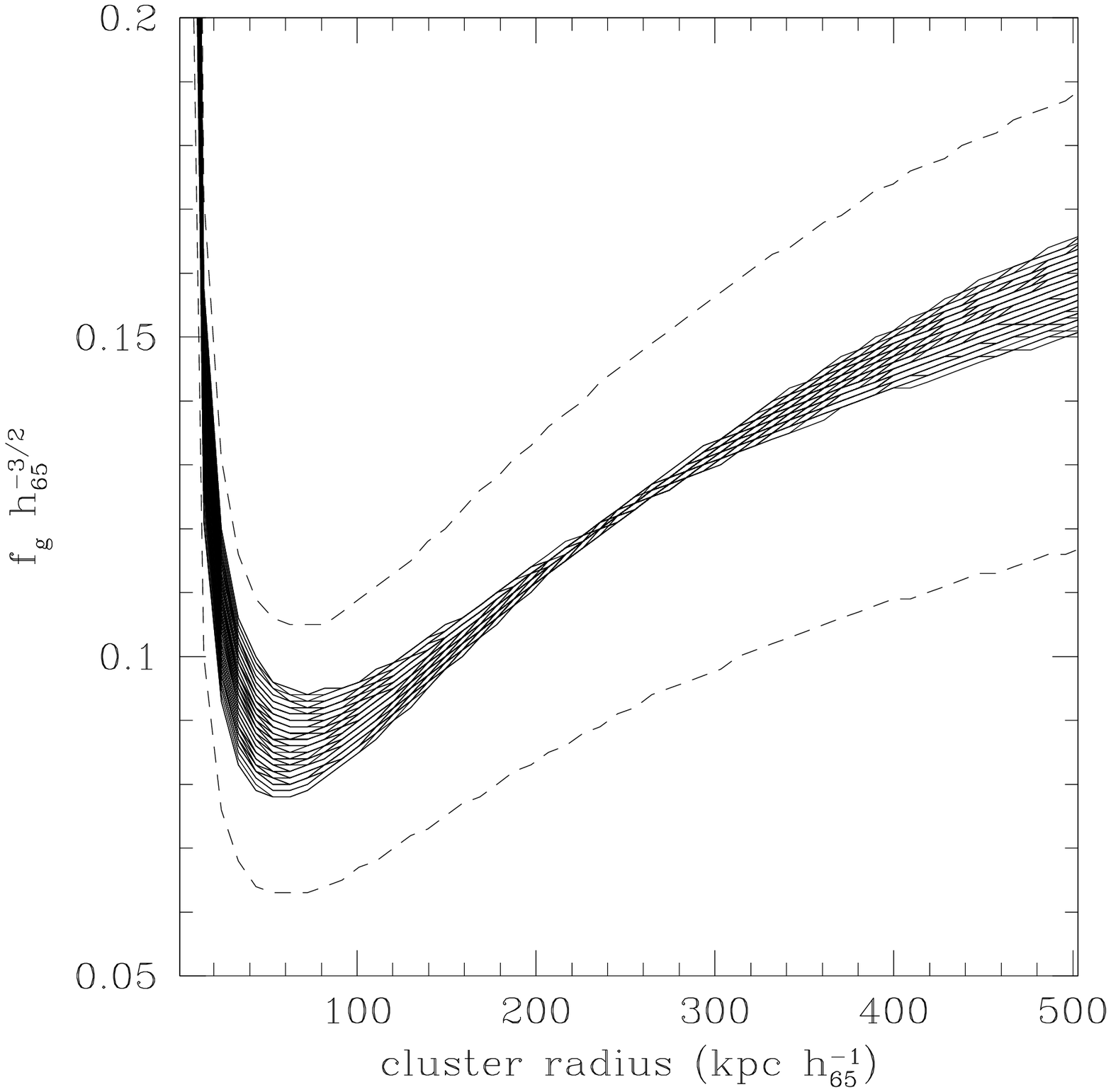}
%\plottwo{ne_vs_rkpc_h65.ps}{fg_vs_rkpc_h65.ps}
\caption{Left. Electron number density versus radius in kpc.  Right. 
The gas mass fraction of the cluster as a function of cluster radius
in kpc.  The solid lines show n$_e(r)$ and $f_g(r)$ for parameter grid
points within the 90\% confidence region of the fitted model.  The
dashed lines show the full 90\% confidence intervals for n$_e(r)$ and
$f_g(r)$, which includes the uncertainty in the spectrum's
normalization for n$_e(r)$, and in the spectrum's normalization and
emission-weighted $T_e$ for $f_g(r)$.  The calculations are done using values \omegam\ = 0.3
and \omegal = 0.7.}
\label{fig:nefgvsr}
\end{figure}

The resulting gas mass fraction profile is shown in
Figure~\ref{fig:nefgvsr}, where again we use solid lines to trace 
$f_g(r)$ for
the allowed shape parameters and the dashed lines to delineate the 90\% confidence 
interval
which includes the uncertainties in $N_{XS}$ and $T_e$.  We compare
the left and right panels of Figure~\ref{fig:nefgvsr} and note that the
statistical uncertainty is dominated by the uncertainty in $T_e$.

We compare the gas mass fraction we calculate with the \chandra\ data
with the gas mass fraction determined using Sunyaev-Zel'dovich effect
data.  We use the method described in \cite{grego2001} which was used
to determine
the gas mass fraction from interferometric SZE data and published
\asca\ temperatures. The method uses the SZE data to constrain
the cluster's shape parameters \be\ and \rc\ and X-ray spectra to 
determine the
emission-weighted temperature, which is combined with the SZE
normalization and shape parameters to determine $n_{e0}$.  To better 
compare the
masses, we rederive the SZE $f_g$ using the \chandra\ temperature
from Section \ref{subsec:temperature}.  
% \cite{grego2000b} found a gas mass within 65$''$ of
% 1.6$^{+0.6}_{-1.2} h^{-2} \times 10^{13} M_\circ$ and a gas mass 
%fraction of 0.062$^{+0.037}_{-0.048} h^{-1}$.
The SZE/\chandra-determined gas mass within 63$''$ is 
$3.0^{+0.7}_{-1.0}
 \times 10^{13} h_{65}^{-2}$ M$_\odot$ and the gas mass fraction within 
63$''$
is $0.06^{+0.07}_{-0.03} h^{-1}_{65}$.  The precision of
the temperature measurement is greatly improved from that used in \cite{grego2001}; this is masked
somewhat by our choice here to use the 90\% confidence interval rather
than the 68\% confidence interval.  
We compare these to the gas mass and gas mass fraction determined in
 this work from the \chandra\ data alone: a gas mass of
2.1$^{+0.1}_{-0.1} \times 10^{13}h_{65}^{-5/2}$ M$_\odot$ and a gas
mass fraction of 0.157$^{+0.022}_{-0.044}h_{65}^{-3/2}$.  The results
are summarized in Table~\ref{table1}.  The gas mass
and gas mass fraction measurements are marginally consistent at
90\% confidence.

To compare the total mass estimate with the weak lensing measurement,
we calculate the total cluster mass projected within a radius of
2$'$.  We integrate the mass density along the line of sight to
$\sim$20 core radii, or about 1.5 Mpc; if we integrate out twice as
far, the total mass changes by only $\sim$15\%. The projected cluster
mass within 2$'$ determined from the \chandra\ data is
6.0$^{+1.7}_{-1.2}\times$10$^{14}$M$_\odot h_{65}^{-1}$.  This is
quite consistent with the minimum projected mass of $(4.9\pm1.6)\times
10^{14} h_{65}^{-1} M_\odot$ calculated from the \cite{clowe2000} weak 
lensing
estimate.

We note here that the cluster's lensing-derived mass profile appears
more compact than expected.  In their weak lensing analysis, \cite{clowe2000} compare the calculated mass
profile to the \cite{navarro1996} universal profile and find \ms1137\ 
is more
centrally concentrated than is usual for massive clusters in this
scenario. (The \cite{navarro1996} profile has since
been noticed to generally predict {\it steeper} density distributions 
than
observed, so this is particularly striking.) \ms1137\ also
exhibits a number of lensed arcs in the cluster center, again
indicating a high central mass concentration.  \cite{clowe2000}
find these points suggestive, but not conclusive, of an elongation of
the cluster along the line of sight.
Our analysis of this cluster shows a core radius of $\sim$ 100 $_{65}^{-1}$ kpc.
This is certainly more compact than the bulk of nearby clusters, whose
core radii are generally around 150 $h_{65}^{-1}$ kpc \citep{jones1999,mohr1995}.  
Only
11 of the 45 clusters in the \cite{mohr1995} sample have core radii as 
small or
smaller than this, yet the temperature of \ms1137\ has a higher
emission-weighted temperature, and is therefore presumably more 
massive, than 28 of the clusters in the sample.  
The cluster may then fall into the compact-core radius group of
\cite{ota2002}, which divides clusters into two distinct groups, one
with core radii around 60 kpc, and one with core radii 220 kpc.
Although a full investigation into correlations between core radius 
size and other properties has yet
to be done, \cite{ota2002} find evidence that clusters with
cD galaxies have small core radii and that the luminosity-temperature
relationships for the two cluster populations have different
normalizations, though similar slopes.

%fg 0.099   0.059   0.082
%mgas 7.390    6.759    7.050

% age of universe in Gyr as fct of cosmo parameters
%  0.30  0.70  0.30   16.31
%  0.30  0.50  0.65    7.00
%  0.30  0.70  0.65    7.53
%  0.30  0.90  0.65    8.22
%  0.30  0.70  1.00    4.89
%  0.20  0.80  0.65    8.88
%  1.00  0.00  0.65    4.40

\section{Cluster Distance}
\label{sec:dist}
We derive the cluster's angular diameter distance by comparing the
cluster's X-ray emission and Sunyaev-Zel'dovich Effect (SZE) data.
The method closely follows the method used in \cite{reese2000a},
except that in this work we use \chandra\ imaging and spectral data
instead of \rosat\ imaging and \asca\ spectral data.

The SZE has a magnitude proportional to the Compton $y$-parameter, \ie, 
the total
number of scatterers, weighted by their associated temperature, 
\begin{equation} y = {{k \sigma_T}\over {m_e c^2}} \int {n_e(l) T_e(l) 
dl}, \label{eq:compton}\end{equation}
where $k$ is Boltzmann's constant,  $\sigma_T$ is the Thomson
scattering cross section, $m_e$ is the electron mass.  For an
isothermal beta-model, the projected SZE effect would be:
\begin{equation}
\Delta T = f_{(x, T_e)} T_{CMB} D_A \! \int\!\! d\zeta \, \sigma_T n_e
\frac{k_B T_e}{m_e c^2} = \Delta T_0 \left ( 1 +
        \frac{\theta^2}{\theta_c^2} \right )^{(1-3\beta)/2},
        \label{eq:szsignal}
\end{equation}
where $\Delta T$ is the thermodynamic SZE temperature
decrement/increment and $f_{(x, T_e)}$ is the frequency dependence of 
the
SZE, with $x=h\nu/kT_{CMB}$, where $T_{CMB}$ is the temperature of the
cosmic microwave radiation $T_{CMB}$=2.728
\citep{fixsen1996}.  In the non-relativistic and Rayleigh-Jeans limits,
$f_{(x, T_e)}\sim -2$; we apply the relativistic corrections of
\cite{itoh1998} to fifth order in $kT_e/m_ec^2$, so that $f_{(x,
T_e)}= -1.907$.

Extending Equation~\ref{eq:betamodel}, the cluster's projected X-ray surface brightness is \begin{equation}
S_X = \frac{1}{4\pi (1+z)^4} D_A \! \int \!\! d\zeta \, n_e n_H 
\Lambda_{eH}
        \:\;\! = S_{X0} \left ( 1 + \frac{\theta^2}{\theta_c^2} \right
        )^{\frac{1}{2}-3\beta},  \label{eq:xsignal}
\end{equation}
where $S_X$ is the X-ray surface brightness in erg s$^{-1}$
cm$^{-2}$ arcmin$^{-2}$, $z$ is the cluster's redshift, n$_H$ is
the hydrogen number density of the ICM, $\Lambda_{eH} = 
\Lambda_{eH}(T_e,
\mbox{abundance})$ is the X-ray cooling function of the ICM in the
cluster rest frame in erg cm$^3$ s$^{-1}$ integrated over
the redshifted \chandra\ band, and $S_{X0}$ is the X-ray surface 
brightness at the center of the cluster.  We convert the X-ray
detector counts to cgs units using the \chandra\ response matrices
available from http://asc.harvard.edu/cal/Links/Acis/acis/; the
conversion factor is 7.91$\times 10^{-12}$ erg cm$^{-2}$ s$^{-1}$.
One can solve for the angular diameter distance by eliminating 
$n_{e0}$. 

We perform a joint fit to the interferometric SZE data and the 
\chandra\ image, in which the X-ray pointlike sources have been masked
out.  We compare the X-ray data to a beta model with a constant cosmic
background; this model is exposure-corrected before the fit.    The logarithm of the
Poisson likelihood is then calculated. The interferometric SZE 
observations provide
constraints in the Fourier (\uv) plane, so we perform our model fitting 
in the \uv\
plane, where the noise properties of the data and the spatial
filtering of the interferometer are well defined. The Gaussian
likelihood is calculated using the SZE model and the SZE data. There 
are no
detected point sources in the SZE data.
We use a downhill simplex method \citep{press1992} to search the
parameter space and maximize the joint likelihood of the cluster position, $\beta$,
$\theta_c$, $S_{X0}$, $\Delta To$, and cosmic X-ray 
background.

Each data set is independent, and likelihoods from each data set can
simply be multiplied together to construct the joint likelihood.
From this we can generate confidence regions for the parameters. 

Uncertainties in the angular diameter distance from the fit parameters
are calculated by gridding (stepping) in the interesting parameters to
explore the $\Delta S$ likelihood space.  As a compromise between 
precision and computation
time, we grid in $\Delta T_0$, $S_{X0}$, $\beta$, and $\theta_c$
allowing the X-ray background to float while fixing
the positions of the cluster (both SZE and X-ray).  
From this four dimensional $\Delta S$ hyper-surface,
we construct the 90\% confidence interval for $D_A$ due to $S_{X0}$, $\Delta T_0$, $\beta$, and
$\theta_c$ jointly.  The correlations between the beta-model
parameters require this treatment to determine accurately the
uncertainty in $D_A$ from the fitted parameters.  We emphasize that these
uncertainties are meaningful only within the context of the spherical
isothermal $\beta$ model.

In the \cite{reese2002} work, the calculated angular diameter 
distance of \ms1137\ using
\rosat\ and \asca\ data was 3179$^{+1103}_{-1640}h^{-1}$ Mpc
(statistical uncertainty only), at 68\%
confidence.  In our analysis
here, with a more accurate emission-weighted temperature and improved
X-ray imaging data, we find an angular diameter distance (at 90\% confidence) of
3439$^{+1854}_{-2087}h^{-1}$ Mpc, which implies a Hubble constant of
31$^{+19}_{-17}$ km s$^{-1}$ Mpc$^{-1}$ (with statistical
uncertainties only).  The \chandra\
observation moderately improves the precision of the measurement, but 
do not rectify it with the \cite{reese2002} sample; the derived Hubble
constant is still $\sim 2.5 \sigma$ from the 18 cluster sample average..  \ms1137\ is a peculiar cluster, and in
the next section we discuss our calculations and attempt to discern
how this cluster may deviate from a relaxed, spherical, isothermal 
cluster.

\section{Discussion} 
\label{sec:discussion}
The X-ray and optical images of the distant galaxy cluster
MS1137.5+6625 show the X-ray surface brightness and optical light have
a regular, symmetric, and quite compact distribution.  However, 
several
lines of evidence suggest that the cluster structure is more complicated:
\begin{itemize}
\item The hardness ratio map and spectral analysis suggest that the
cluster has a relatively more complicated temperature and metal
abundance structure than density structure.
\item The gas mass fraction calculated from the SZE/\chandra\ data is
only marginally consistent with the gas mass fraction calculated from
the X-ray data.
\item The Hubble constant calculated for this cluster is low by a
factor of $\sim$2. (This is closely related to the discrepancy in the
gas mass fraction and gas mass.)
\end{itemize}
These calculated quantities are summarized in Table~\ref{table1}.  The
most compelling argument that \ms1137\ deviates from isothermal,
spherical symmetry is that the Hubble constant value derived from it
is so low.  The X-ray and SZE data sets are not sensitive enough
individually to allow us to definitively test our assumptions, but
taken together they do provide some insight.  Here, we discuss a
number of possible explanations for our observations.
\subsection{SZE Data Systematics}
The possible contribution to systematic uncertainty to the Hubble
constant and gas mass fraction measurement by the SZE data
itself is very low.  It is possible that there are undetected
radio-bright point sources in the cluster field.  Unless they are
unfortunately placed on the sidelobes, point sources will decrease the
measured SZE effect and increase the measured \Ho.  Combining the SZE
observations with 1.4 GHz NVSS \citep{condon1998} observations
suggests that possible undetected point sources in the \ms1137\ field
have a negligible effect on the derived Hubble parameter
\citep{reese2002}.  We find no evidence for a radio halo
in this cluster in the NVSS image.  Radio haloes, \ie, large scale
diffuse radio emission, would again depress the measured SZE and
therefore would cause an overestimate of the Hubble constant.

Anisotropies of the CMB of other types which might affect the SZE data
are at very low levels.  A limit to the primary anisotropies at these
angular scales was set by \cite{dawson2001} and \cite{holzapfel2000b}.  
They find the
Rayleigh-Jeans temperature of these fluctuations to be less than 14
$\mu$K, and this in turn should contribute less than 2\% uncertainty
to \Ho.  Confusion from the kinematic SZE effect (the spectral
distortion of the CMB made by the cluster moving with respect to the
CMB rest frame) should also be quite small; for an 8 keV cluster moving
with a 300 km s$^{-1}$ velocity component in the line of sight
direction, the kinetic SZE effect will be $\sim$4\% of the thermal SZE (and
the sign of the effect will depend on whether the cluster is moving 
towards or away
from the observer) leading to an 8\% error in the Hubble constant.
It is therefore quite unlikely for the SZE to be severely in error due
to peculiar velocity of the cluster gas.

\subsection{Geometry}
The Hubble constant is derived under the assumption that the cluster
can be characterized by a spherically symmetric distribution.  This
will be in error by the ratio of the projected cluster size $L_{PROJ}$ 
to the
line of sight cluster size $L_{LOS}$: $H_{meas}=H_{true}\times\,
L_{PROJ}/L_{LOS}$.  We have
measured the apparent axis ratio to be 0.87$^{+0.06}_{-0.07}$ for the
elliptical beta model; the core radius measured by the
circular beta model is roughly the geometric mean of the two axes.  For
 an apparently circular distribution, the line of
sight axis could be longer or shorter than the projected axis, leading
to a prolate or oblate three dimensional cluster, respectively.  
For an apparently elliptical mass distribution, the true three-dimensional
distribution has another degree of complexity in addition to the
degree of prolateness or oblateness; the axis
of symmetry could have any angle of inclination with respect to the
plane of the sky, allowing the symmetry axis an essentially infinite 
range of lengths which would produce the same projected image.  
Additionally, there is no
a priori reason that clusters need to be ellipsoidally symmetric; in
fact, for axis ratios of less than $1/\sqrt{2}$, the concentrically 
ellipsoidal isothermal beta-model for the density distribution is unphysical
\citep[\cf][]{grego2000a} if the the cluster is indeed in hydrostatic equilibrium.  The gravitational potential supporting such a
beta-model requires {\emph {negative}} dark matter density.  A more 
sophisticated model for the non-spherical density distribution is
clearly required in such cases, but is beyond the scope of this paper.  
We note that for measuring cosmological parameters from an ensemble of clusters,
increasing the complexity of the model is not strictly necessary. For
example, \citep{sulkanen1999} shows that from a sample of triaxial
clusters, one can reconstruct an unbiased estimator for the Hubble
constant by using a spherical model.  

In the simplest picture, the error in the Hubble constant is approximately the  ratio of the apparent cluster size (\ie, the
geometric mean of the two core radii) to the line of sight size.  In
such a picture, for
\ms1137's measured Hubble constant to equal the sample mean of
\cite{reese2002}, L(projected)/L(line of sight) must be 0.58.  As a
fiducial point, in a flux limited sample of clusters fitted with
elliptical beta-models, \cite{mohr1995} measure an axis ratio for the
sample of 0.8$^{+0.14}_{-0.11}$.  For the low Hubble constant to be
due solely to an elongated cluster, it would have to be have two 
unusual characteristics for a cluster chosen from an unbiased sample: it would have a much longer axis ratio than generally observed in projection, and would have its symmetry axis nearly aligned
with the line of sight, basically a long cigar pointed at the
observer.  If the symmetry axis were not nearly along the line 
of sight, the true axis ratio would need to be even greater.  

Since such an elongation along the line of sight is so unexpected, this
may indicate that the criteria by which \ms1137\ was detected favored such
clusters.  \ms1137\ was originally identified in the EMSS \citep{gioia1990}, using a
2.$'4\times2.'4$ detection cell, a size which would just fit the
cluster emission we observe with \chandra.  It was detected with a
signal-to-noise ratio of 5.3 in a survey with minimum S/N for
detection of 4.0.  \ms1137\ is the coolest
and least luminous of the z $>$ 0.5 clusters in the EMSS survey.  If
there does exist a population of similarly luminous, highly
elongated clusters at high redshifts, it is likely that only those
aligned very nearly along the line of sight would be detected by this
survey; a cluster also near the detection cutoff but which was elongated in the plane of the
sky would deposit less of its flux into a single detection cell and
may not be detected.  For example, if \ms1137\ had an intrinsic axis
ratio of $\sim$0.75 but with the major axis in the plane of the sky,
it is unlikely that it would have been detected with sufficient
significance to be included in the survey.  

The selection effects from the original point-source algorithm used
for the EMSS were discussed in \cite{pesce1990}, and reprocessing of
the Einstein data led \cite{lewis2002} to conclude that the EMSS survey
systematically missed clusters of low surface brightness.  Similarly, simulations
of the EMSS detection algorithm by \cite{ebeling2000} suggest the
algorithm may miss clusters with significant substructure.

\subsection{Temperature Gradients}
Although we have no compelling evidence for a cooling flow at the cluster's center, we do have
evidence of structure in the temperature and abundance distribution.
Temperature structure could affect both the derived shape parameters
and the derived emission-weighted temperature.

Temperature structure will affect the X-ray and SZE data differently.
It should not strongly affect the X-ray-derived spatial shape
parameters; for a given emission measure, the emitted flux in the band
we use for spatial fitting only changes by $\sim$ 20\% for gas with
kT$_e$ from 4 keV to 10 keV. Also, for the X-ray spatial fitting, the data are
azimuthally averaged, so that such structure would have little effect.

The SZE measurements, by virtue of the spatial filtering inherent in
interferometry, are insensitive to temperature gradients on scales 
larger than a few arcminutes, but can be quite sensitive to smaller
scale gradients.  As the SZE effect is proportional to $n_e \times
T_e$, the density inferred at any particular point will be over
(under) estimated from an under (over) estimate of the temperature.  
Temperature structure could result in the beta-model fitted to the SZE differing from
that fitted to the X-ray data. We do not expect this poor fit to the beta-model 
to strongly affect the calculated Hubble constant, because the joint-fit
to the model is dominated by the X-ray imaging data in this case.  It may, 
however, affect
strongly the gas mass fraction determined from the SZE data, as this
is derived from the SZE spatial fits.  

If \ms1137\ had a non-hydrostatic pressure structure, then we may
expect to see
evidence of this in the SZE data.  The low surface
brightness of the SZE-effect makes it difficult to image on small
scales, however, and with the current data, we are unable to make a conclusion
about non-hydrostatic pressure.  Future generations of SZE instruments
may have such capability.  

The temperature structure of the cluster could contribute to the low
value of the Hubble constant in another way.  The temperature we use 
for the \Ho\
calculation is the emission-weighted temperature, and the relevant
quantity for the calculation is actually the density-weighted temperature.  As
a fiducial point for the discussion, we compare the
emission-weighted temperature and the density-weighted temperature for
a cluster with the following structure: \be\ equal to the best-fit value from our model fit,
and gas within two core radii ($\sim$200 kpc) at half the temperature of the gas from
two core radii to ten: $T_e(r < 2 r_c) = 0.5\times T_e(2 r_c < r < 10
r_c)$. 
Temperature structure of this magnitude, while unusually pronounced,
has been observed by \chandra\ in the cluster Abell 1835
\citep{schmidt2001}, a massive cooling-flow cluster with highly luminous optical emission-line nebulae.  In such a scheme, the emission-weighted
temperature would be 0.80 times the density-weighted temperature, and
the Hubble constant would be 0.64 times the true Hubble constant.  For the underestimation of the
Hubble constant to be due {\em solely} to an underestimate of the 
temperature, the emission-weighted temperature must be $\sim$0.65-0.70
times the density-weighted temperature.  We stress that this is a simple calculation; the
exact effect depends also on the specifics of the instruments used to
make the measurements. However, even if \ms1137\ had as extreme a temperature gradient as the atypical Abell 1835, this could not explain the entire Hubble constant underestimate. 
\section{Summary}
\label{subsec:summary}
In conclusion, the massive, distant galaxy cluster \ms1137\ appears 
compact and regular in its optical and X-ray images, but evidence suggests that it is not an isothermal, spherical cluster.  The
Hubble constant calculated for the cluster from X-ray imaging and
spectra and interferometric SZE data is unusually low.

Many of the known systematic effects in the X-ray/SZE Hubble constant
measurement tend to make the Hubble constant higher, and few effects
make it lower.  (A thorough discussion of these systematic effects can
be found in \cite{reese2002}.)  Two effects which {\em can} make the
measured Hubble constant artificially low are elongation along the line
of sight and temperature structure in the ICM.  We have evidence
suggestive of a line of sight elongation.  The
optical lensing data indicate the surface mass density is quite high, as evidenced by
strongly lensed arcs, and that the mass distribution is very compact.
The X-ray data show the ICM density distribution is unusually compact,
as well.  We see structure in the hardness ratio map which could be
interpreted as temperature structure, with cooler or less
metal-enriched gas at the center of the cluster.

We also compare other calculated quantities besides the Hubble
constant to investigate the systematic effects.  The SZE-derived gas mass is completely insensitive to
geometry, though it is quite sensitive to temperature
structure; if the emission-weighted temperature were biased low, the
SZE gas mass would be an overestimate of the true gas mass.  The
X-ray-derived gas mass is essentially insensitive to temperature
structure, but has a moderate dependence on geometry; if the cluster
were elongated along the line of sight, the X-ray gas mass would be a
slight overestimate.
The X-ray-derived gas mass is 2.1$^{+0.1}_{-0.1}\
h_{65}^{-5/2}\times 10^{13} M_\circ$  and
the SZE-derived mass is $3.0^{+1.0}_{-1.0}\ h_{65}^{-2}\times 10^{13}
M_\circ$.  Within the uncertainty the masses agree
and we cannot place strong limits on temperature structure and
elongation with the current datasets.

We conclude with two points: one, that cluster appearance is a poor
criterion for choosing sources for cosmological tests.   \ms1137\
appeared as a possibly ideal cluster for cosmology, with its regular, compact,
round, X-ray and optical images, and no evidence of a strong cooling
flow.  However, it yields a Hubble constant measurement which is
strikingly low.  We stress that a cluster sample must be selected via
well-defined, objective criteria.  And two, that selection effects in the parent sample from which the
clusters are drawn should be well-understood, \ie, although for cosmological work we generally
select clusters from existing X-ray surveys on the basis of luminosity
rather than surface brightness, any selection for surface brightness
in the X-ray survey itself will still be present in the
luminosity-selected sample.  This situation can be ameliorated by
future cluster surveys, such as the SZE survey discussed in
\cite{holder2001b}, which have selection functions different from the
X-ray and optical cluster surveys.
\section{Acknowledgements}
We are very sad to note that Leon VanSpeybroeck passed away at the end of 2002.
He was an excellent scientist and colleague and friend.  We remain
inspired by his example.

Many thanks to the referee for her insightful comments that improved
this paper.
We gratefully acknowledge the contributions of several people who
developed public-use tools which we used extensively: Maxim Markevitch for the blank field background data sets,
Alexey Vikhlinin for the X-ray data analysis software, and J.\ S.\
Sanders and A.\ C.\ Fabian for the adaptive binning software. We thank the CLEGS
group at the CfA for helpful discussion.  EDR acknowledges support
from the NASA Chandra Postdoctoral Fellowship PF 1-20020.  This work was supported 
in part by the NASA grants NAG5-4985, GO0-1037X, NAG5-10071, and GO1-2138X.

\clearpage
\bibliographystyle{apj}
\bibliography{clusters,apjourn}

\clearpage
\begin{deluxetable}{lcccc}
\setlength{\tabcolsep}{1.5mm}
\renewcommand{\arraystretch}{1.60}
\tablewidth{0pc}
\tablecaption{Summary of Derived Properties\label{table:results}}
\tablehead{
\colhead{Data} & \colhead{gas mass} & \colhead{total mass} & 
\colhead{$f_g$}&\colhead{$D_A$}\\
\colhead{}& \colhead{$10^{13} M_\circ$} &\colhead{$10^{14} M_\odot$} & 
\colhead{}&\colhead{(Mpc)}\\
\colhead{}&\colhead{within 63$''$}&\colhead{projected within 120$''$}& 
\colhead{within
63$''$}& \colhead{(for $\Omega_M=0.3$, $\Omega_\Lambda=0.7$)}
}
\startdata
Chandra  &2.1$^{+0.1}_{-0.1}\ h_{65}^{-5/2}$ &
6.0$^{+1.7}_{-1.2}\ h_{65}^{-1}$ & 0.157$^{+0.022}_{-0.044}\ 
h_{65}^{-3/2}$ & ---\\
SZE  & $3.0^{+1.0}_{-1.0}\ h_{65}^{-2}$ &--- & $0.06^{+0.07}_{-0.03}\
h^{-1}_{65}$  &---\\
SZE/Chandra &--- &--- & --- & 3439$^{+1854}_{-2087}h_{65}^{-1}$\\
Keck Weak Lensing & --- & 4.9$^{+1.6}_{-1.6}\ h_{65}^{-1}$ & --- &---\\
\enddata
\label{table1}
\end{deluxetable}

\clearpage

%\begin{figure}[hbt]
%\epsscale{1}
%\centerline{\hbox{
%\psfig{figure=hardness_sz_color.ps,height=4.0in}
%\psfig{figure=hardness_sz_grey.ps,height=4.0in}
%}}
%\caption{The SZE data contours over the hardness ratio map of
%Figure~\ref{fig:hardness}.  The SZE data have no \uv\ filter
%and represent a compromise between angular resolution and surface
%brightness sensitivity.  The FWHM approximation to the restoring beam
%is in the lower left corner, and is 27$''\times32''$.  The contours
%are negative value and approximately 1.5$\sigma$, or 0.1 mJy/beam.}
%\label{fig:sz_hardness}
%\end{figure}

\end{document}